\documentstyle[epsf,psfig,natbib_mod]{mn}
\begin{document}
\title[SAX~J1808.4-3658 and X-ray variability in X-ray binaries and AGN]
{SAX~J1808.4-3658 and the
origin of X-ray variability in X-ray binaries and active galactic nuclei}
\author[Philip Uttley]
  {Philip Uttley\thanks{E-mail: pu@astro.soton.ac.uk} \\
  School of Physics and Astronomy, University of Southampton,
Southampton SO17 1BJ \\}

\maketitle
\parindent 18pt

\begin{abstract}
The aperiodic X-ray variability in neutron star and black hole X-ray
binaries (XRBs), and active galactic nuclei (AGN) shows a
characteristic linear relationship between rms amplitude and flux, implying a
multiplying-together or `coupling' of variability on different time-scales.  Such a
coupling may result from avalanches of flares, due to magnetic reconnection in
an X-ray emitting corona.  Alternatively this coupling may arise
directly from the coupling of
perturbations in the accretion flow, which propagate to the inner
emitting regions and so modulate the X-ray emission.  Here, we
demonstrate explicitly that the component of aperiodic variability which
carries the rms-flux relation in the accreting millisecond pulsar 
SAX~J1808.4-3658 is also coupled to the 401~Hz pulsation in this
source.  This result implies that the rms-flux relation in
SAX~J1808.4-3658 is produced in the accretion flow on to the magnetic
caps of the neutron star, and not in a corona.  By extension we infer that
propagating perturbations in the accretion flow, and not coronal flares, are the
source of the rms-flux relations and hence the aperiodic variability
in other XRBs and AGN.
\end{abstract}


\section{Introduction}
Aperiodic X-ray variability, occuring over a broad range of time-scales
in the form of `flicker noise' or `band-limited noise' 
is a ubiquitous characteristic of
X-ray binary systems \citep{vdk95} and active galaxies
(\citealt{mch88,utt02,mar03}).  The origin of this variability is not
well understood.  Recently, a new fundamental property of
the aperiodic variability in XRBs and AGN was discovered: the rms-flux relation
\citep{utt01}.   Specifically, there exists a strong linear correlation between
the amplitude of X-ray 
variability on short time-scales and the X-ray flux as measured on
longer time-scales (\citealt{utt01,ede02,vau03}).  Since the flux variations on 
all the time-scales probed in these studies 
are dominated by the aperiodic variability, this linear `rms-flux relation'  
implies that aperiodic variations on longer time-scales modulate the aperiodic 
variations on shorter time-scales, i.e. the aperiodic variations 
on different time-scales are coupled together.  

Importantly, the observed linear rms-flux relations, which occur over a broad range of time-scales, 
cannot be explained if the variability is produced by
simple shot-noise models where the light curve is produced by the summation
of randomly occuring shots or flares which are
independent of one another (see \citealt{utt01} for a detailed discussion).
Instead the shots must be coupled together. 
For example, the shots may correspond to fractal structures, e.g.
flares that break into smaller structures, which in turn break up and so on, 
producing a coupling between larger slower variations and smaller faster ones. 
Alternatively, the required coupling may be produced if flares
trigger avalanches of smaller flares (e.g. similar to the model of \citealt{ste96}).
A natural interpretation of such models is that they correspond to flares due to
magnetic reconnection in the corona \citep{pou99}, so we refer to this class of
explanations for the variability as the `coronal-flare' or CF model.

Alternatively, a linear rms-flux relation can be naturally produced by the model
of \citet{lyu97}, where the variability is caused by variations in accretion rate 
occuring at different radii (with slower variations occuring at larger radii),
which propagate inwards and modulate the X-ray emitting region
and hence the X-ray
light curve.  In this model, the amplitudes of accretion rate variations produced 
at a given radius scale with the local accretion rate at that radius (which is
driven by the slower variations that originated at larger radii) and hence variations
on different time-scales are coupled together.  In fact, any model of
propagating perturbations in the accretion flow (e.g. density waves, \citealt{mis00})
could produce a linear rms-flux relation provided the perturbations can couple 
linearly together.  We will call this class of explanations for the variability
the `propagating perturbation' or PP model.

In this Letter, we point out that the CF and PP models can be distinguished in 
the case of accreting neutron star XRBs, which show a linear rms-flux relation in their
aperiodic X-ray variability, provided that it can be shown that the aperiodic
X-ray variability containing the rms-flux relation originates on or close to
the neutron star surface, and hence cannot be associated with coronal flares. 
We demonstrate explicitly that the accreting millisecond X-ray pulsar SAX~J1808.4-3658
fulfils these criteria, so that the CF model can be ruled out in this case and the
PP model is the most likely explanation for the aperiodic variability
and the rms-flux relation in this system. 
We then infer that the PP model is also the correct explanation for 
the aperiodic variability which shows a linear rms-flux relation in other XRBs,
and AGN, before discussing the further implications of this result.

\section{The origin of the rms-flux relation in SAX~J1808.4-3658}
In \citet{utt01} we demonstrated that the linear rms-flux relation is observed 
in the aperiodic X-ray variability of the accreting millisecond X-ray pulsar
SAX~J1808.4-3658.  Due to its highly coherent 401~Hz X-ray pulsations \citep{wij98a}
SAX~J1808.4-3658 is a confirmed NS system, and offers the possibility of directly
testing the CF and PP models for aperiodic X-ray variability, if it can be shown 
that the X-ray variability originates on or close to the neutron star surface.  This is
because although the CF model requires that the X-ray variability originates in a 
flaring corona, the PP model does not require a specific physical origin for the
X-ray emission, provided that the perturbations in the accretion flow can
propagate in to the X-ray emitting region (whether this is an accretion powered
corona or the surface of a neutron star) and so modulate the X-ray emission.

\subsection{Coupling of the 401~Hz pulse and aperiodic flux variations}
\label{pulsermssec}
Evidence that much of the X-ray emission in neutron star XRBs originates at the 
neutron star surface (or some boundary layer close to the surface)
comes from comparisons of X-ray spectra of neutron star and black hole
candidate systems \citep{don03}.  Fourier-frequency resolved
analysis of neutron star XRB spectral variability confirms this result
and further suggests that some component of the variability originates
at the neutron star surface or boundary layer, provided certain
plausible assumptions are made
about the spectral shape of emission from that surface \citep{gil03}.
However, we require more model-independent evidence
that the aperiodic variability in SAX~J1808.4-3658 originates at or close
to this surface.  Interestingly, such evidence is provided by
a coupling between the aperiodic variability and the 401~Hz pulsation in this
system, which has been inferred from a broadening of the pulse peak measured in the
power spectrum of this source \citep{men03}. 
This result implies that a significant component of
the aperiodic X-ray variations originates on the neutron star surface,
at the accretion-fed magnetic caps, and is modulated at the spin frequency
to produce the observed coupling. 

The method used by \citet{men03} to demonstrate coupling between the
aperiodic and periodic variability in SAX~J1808.4-3658
tests the assumption that the observed power spectrum results from a
simple convolution of the Fourier transforms of aperiodic and periodic
variations.  In the time domain, this assumption corresponds to the case where the
light curve of aperiodic variations is multiplied by a sinusoidal 
pulse light curve with a mean of unity and an rms equal to the fractional rms of
the pulse.  Then it is easy to see that a
prediction of the coupling of the pulse and aperiodic variability
reported by \citet{men03} is that the {\it absolute} rms of the pulse
should scale linearly with the aperiodic flux variations, i.e. the pulse will
itself show a linear rms-flux relation.
\begin{figure}
 \par\centerline{\psfig{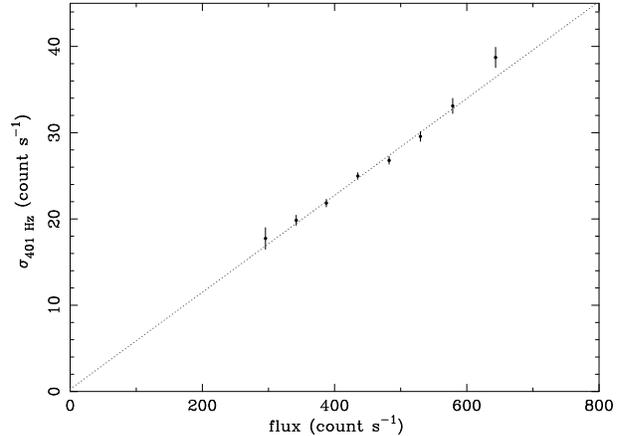}}
 \caption{\label{pulserms}
   The rms-flux relation of the 401~Hz pulsation in SAX~J1808.4-3658.
Standard ($1\sigma$) errors are determined from the spread of the individual
401~Hz power measurements binned together in the averaged PSDs.  The
dotted line shows the linear best-fitting model described in the text.
   }
\end{figure}

To test this prediction we used data from a {\it Rossi X-ray Timing
Explorer} (RXTE)
observation of SAX~J1808.4-3658 used by \citet{utt01} and
\citet{men03}\footnote{Observation IDs 30411-01-06-000, 30411-01-06-00
observed 1998 April 18}.
A 2-20~keV light curve (with $2^{-11}$~s resolution), of 23~ks exposure was
extracted from the data and split into segments of 1~s length.  The power spectrum
and flux (count rate) was measured for each individual segment, and the
power spectra of many segments were averaged together according to their
flux (here, power spectra are measured in
$rms^{2}$ units, i.e. not normalised by the square of the mean flux).
We estimated the noise level directly from the data
by fitting a power-law to the surrounding 200~Hz window of data, centred on but
excluding the peak at 401~Hz,
and took the square-root of the noise-subtracted power at 401~Hz
(directly equivalent to the integrated power or variance, since the frequency
resolution is 1~Hz) to obtain the absolute rms of the pulse as a function of flux.  The resulting 
rms-flux relation is shown in Fig.~\ref{pulserms}.  A linear fit to the data
provides a good fit ($\chi^{2}=8$ for 6 degrees of freedom), and 
yields an offset on the flux axis consistent with zero 
(flux offset $C=-5\pm38$~count~s$^{-1}$, errors are 90\% confidence).

\subsection{Coupling of the pulse to the aperiodic variability amplitude}
The linear rms-flux relation shown by the pulse confirms that
aperiodic and periodic variations are coupled, and therefore at least
some component of the aperiodic X-ray
variability must originate at or close to the magnetic
caps of the neutron star.  However, this does not necessarily imply
that the component of aperiodic X-ray variability showing a linear rms-flux
relation also originates at the same location.  For example, if a
component with a linear rms-flux relation, which is not coupled to
the pulse, contributes a similar flux to a
component with constant rms which is coupled to the pulse, both the
pulse and aperiodic rms-flux relations can be preserved.  This is
because the highest and lowest total observed fluxes correspond to the times
when the fluxes of {\it both} aperiodic components are high and low
respectively.  Hence, when the total flux is high both the pulse rms
and the aperiodic rms will be high, and the rms of the pulse and
aperiodic variability will both be low when the total flux is low.

However, in the case where the aperiodic rms-flux relation is
produced in a separate component to the pulsed component of aperiodic
variability, we expect that the amplitude of the aperiodic variability and
the rms of the pulse will not be correlated.  To demonstrate this
fact, we use simulated light curves.
 As shown in \citet{utt03} and Uttley, M$^{\rm c}$Hardy \&
Vaughan (in prep.), aperiodic light curves with a linear rms-flux relation on all
time-scales may be simply generated by replacing each data point in a
linear light curve (e.g. generated using the algorithm of
\citealt{tim95}) with its exponential.  We examined two cases:
\begin{enumerate}
\item The aperiodic variability is produced by a single component with
a linear rms-flux relation on all time-scales, which is modulated by a
sine wave at the pulse frequency (i.e. in its simplest form this model
corresponds to the case where the rms-flux relation is
produced at the magnetic caps of the neutron star).
\item The aperiodic variability is produced in equal parts by two components: a
linear component with rms independent of flux (which is modulated by a
sine wave at the pulse frequency), and an unpulsed component with a
linear rms-flux relation on all time-scales (i.e. the rms-flux
relation is produced in a different location to the magnetic caps, e.g. in a corona).
\end{enumerate}
In both cases, the simulated aperiodic variability assumed a broken
power-law power spectrum with power-law slope 0 below 0.3~Hz and -1 at higher
frequencies (up to the Nyquist frequency), to approximate the observed
power spectrum \citep{wij98b}.
The simulated amplitudes of variability (combined amplitude in
case~(ii)) were chosen to match that of the source, and Poisson noise
corresponding to the observed source plus background count rates
was also included.
\begin{figure}
 \par\centerline{\psfig{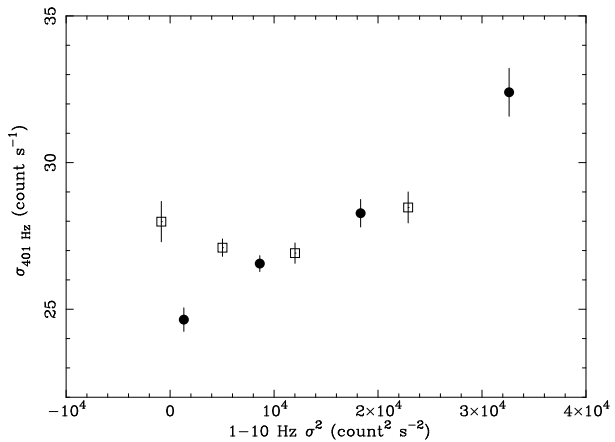}}
 \caption{\label{simrmsvar}
   Pulsation rms versus aperiodic 1-10~Hz variance for simulated light curves
of SAX~J1808.4-3658.  Filled circles denote case (i), where a single
component of aperiodic variability containing the linear rms-flux relation is
modulated by the pulse.  Open squares denote case (ii), where there
are two aperiodic components and the pulse is only coupled to the
component which has rms independent of flux.  Errors are determined
using the same method as for Fig.~\ref{pulserms}.
   }
\end{figure}

To examine the correlation between the amplitude of aperiodic
variability and the pulse amplitude, we averaged together
the 1~s measurements of
the power spectrum (discussed in
Section~\ref{pulsermssec}) according to their noise-subtracted
variance (i.e. integrated
power) measured in the 1-10~Hz band, to obtain the pulse rms as a
function of the aperiodic 1-10~Hz variance.  We bin the pulse rms as a
function of 1-10~Hz variance rather than rms, because in individual
1~s segments the true noise level may exceed the estimated value,
which can lead to negative noise-subtracted variance, so that rms cannot be
determined (e.g. see discussion in \citealt{utt01}, \citealt{gle03}).

Fig.~\ref{simrmsvar} shows the pulse rms versus aperiodic variance for the
case (i) and case (ii) simulations described above.  Case (i), where the pulsed aperiodic
variability contains a linear rms-flux relation, shows a clear
correlation between pulse rms and aperiodic variance, with a linear
plus constant model $\chi^{2}$ fit yielding a positive gradient $(2.3\pm0.4)\times10^{-4}$
for $\chi^{2}=1.3$ for 2 degrees of freedom (errors are 90\%
confidence limits).  Note that the increase in pulse rms with
aperiodic variance is relatively small, because the aperiodic variance
contains a large amount of intrinsic scatter, unrelated to the
rms-flux relation, due to the stochastic nature of the variability and
the effects of noise.  Nonetheless, a significant correlation between
pulse rms and aperiodic variance is easily seen.
Case~(ii), on the other hand, shows no such
correlation (gradient $(0.4\pm0.5)\times10^{-4}$ for $\chi^{2}=5.9$,
for 2 degrees of freedom), due to the fact that the pulsed aperiodic
component does not contain the linear rms-flux relation.  Note that the spread
in 1-10~Hz variance is larger in case~(i) than in case~(ii) because of
the stronger rms-flux relation (which applies to the entire light
curve) in case (i), but the average variance is the same in both cases.
\begin{figure}
 \par\centerline{\psfig{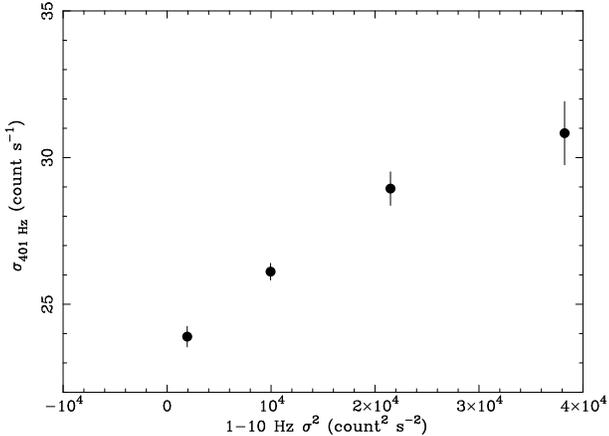}}
 \caption{\label{obsrmsvar}
   Observed pulsation rms versus aperiodic 1-10~Hz variance for
2-20~keV light curve of SAX~J1808.4-3658. Errors are determined
using the same method as for Fig.~\ref{pulserms}.
   }
\end{figure}

In Fig.~\ref{obsrmsvar} we plot the observed pulse rms versus
aperiodic variance.  The pulse rms and the aperiodic variance are
clearly correlated (gradient $(2.2\pm0.4)\times10^{-4}$ for $\chi^{2}=3.2$,
for 2 degrees of freedom).  Furthermore, the observed gradient of the
correlation is consistent with that predicted by the case (i)
simulated data.  Therefore, we conclude that the aperiodic variability
containing the linear rms-flux relation is coupled to the pulse, and
furthermore, no additional component of aperiodic variability is
required by the data.   Thus we infer that the X-ray variability
carrying the rms-flux relation
originates at the magnetic caps of the neutron star and therefore the
PP model is correct in SAX~J1808.4-3658 and the CF model is ruled
out.  For completeness, we note here the possibility that case (i) corresponds to a more
complicated PP model,
where most of the aperiodic variability is produced before the pulsed
variability by perturbations propagating through a separate unpulsed emitting
region (e.g. a static corona above the disk), before reaching the site
of the pulsed emission. 
However, given the available spectral-variability evidence for other neutron
stars \citep{gil03}, it is simplest to assume that all the variable emission originates at
or close to the neutron star surface.

\section{Discussion}
We have demonstrated that the aperiodic X-ray variations which contain the
linear rms-flux relation in SAX~J1808.4-3658 are coupled to the
401~Hz pulsation, and hence originate on or close to the magnetic
caps of the neutron star.  The only existing model which can explain
this result is the PP model: specifically, the rms-flux relation is
produced by the coupling of perturbations in the accretion flow as
they propagate inwards.  When the accretion flow is channeled on to
the magnetic caps of the neutron star, the energy of accretion is
released and the pattern of fluctuations in the accretion flow
(including the rms-flux relation) is imprinted on the
resulting X-ray emission. 

Although we have only shown that the PP
model is the most likely explanation of the rms-flux relation in
SAX~J1808.4-3658, it seems highly likely also that the PP model
explains the aperiodic variability and rms-flux relation in other accreting
neutron star and black hole systems (including AGN).  For example,
the power spectra of broadband noise in both black hole and neutron star XRBs (including
SAX~J1808.4-3658) can be described by a simple model involving the superposition of broad
Lorentzian features (\citealt{bel02,pot03,str03}), and both neutron star and black hole systems
show similar correlations between the
different characteristic frequencies of these features
(e.g. \citealt{wij99,bel02,pot03}).
The strong similarities between black hole XRB variability and that of AGN, with
characteristic time-scales apparently scaling with the black
hole mass (e.g. \citealt{utt02,mar03,mch03}) also strongly support
the idea that AGN and XRBs share the same aperiodic variability
mechanism.  Since the light curves of all these types of
system show a linear rms-flux relation
(e.g. \citealt{utt01,vau03,gle03}), the application of Ockham's
Razor would suggest that the rms-flux relation is produced by the same
mechanism in all cases and hence the aperiodic X-ray variability in
neutron star and black hole XRBs and AGN is produced by a PP
mechanism, and not by coronal flares.

It should be restated here that the PP interpretation of the variability
does not carry any implications for the existence of an X-ray emitting
corona.  This is because the aperiodic variability (and the rms-flux
relation imprinted in it) is produced in the accretion flow and is
independent of where the X-rays are emitted, provided that variations
in the accretion flow can modulate the X-ray emission.  Thus a
corona which is heated by many small reconnection events (too small to have
a very strong effect on the variability) remains a viable source of
the X-ray emission (in black hole systems, at least).  
In fact, by invoking an extended X-ray emitting region (such as a
corona) which possesses
a temperature gradient, so that higher energy X-rays are
preferentially emitted closer to the black hole,
PP-type models naturally produce the time-dependent delays between energy
bands and the energy-dependent shape of the power spectrum, which are
observed in black hole X-ray binaries
(e.g. \citealt{mis00,kot01,zyc03}) and AGN
(e.g. \citealt{vau03,mch03}).

We have recently shown (\citealt{utt03}, and Uttley et al., in prep.) that the
existence of an rms-flux relation on all time-scales leads naturally
to the appearance of non-linear behaviour which is 
observed over a range of time-scales in both XRBs
(\citealt{mac02,gie03}) and AGN (\citealt{utt99,gli02}),  Although the
PP model is not a necessary explanation for such behaviour (which is
a phenomenological outcome of the rms-flux relation), it
provides a natural physical framework for understanding it, in that
the PP model appears to be the correct physical explanation for the
rms-flux relation.  

The PP model allows a simple
understanding of other aspects of the variability.  For example, the
characteristic time-scales observed in the variability of XRBs and AGN
can be understood in terms of time-scales in the accretion flow
(e.g. the broad Lorentzian features can be formed
if perturbations are produced over certain narrow ranges of
radii), thus potentially
unifying models for the aperiodic variability with models for
quasi-periodic oscillations that are sometimes observed in XRBs
(though not yet conclusively in AGN). What physical time-scales these
characteristic features correspond to is not yet clear.  The PP model
requires that perturbations are produced at different radii in the accretion
flow, that these perturbations can propagate inwards without being
suppressed (e.g. damped by viscous dissipation), and that the perturbations
produced at different radii can couple together.  A possible
configuration which could fulfil these criteria is a geometrically
thick accretion flow (such as a thick disk, or an advection dominated
accretion flow), where accretion time-scales are relatively
short so that variations, e.g. on the thermal time-scale, can propagate
to the inner regions of the accretion flow without being significantly
damped (\citealt{man96,chu01}).  Alternatively, the fact that a linear rms-flux
relation is observed in Cyg~X-1 in both the soft and hard states has
been used by \citet{gle03} to suggest that the variations originate in
a coronal accretion flow (rather than a disk), which should also show a
short accretion time-scale and hence might satisfy the PP model
requirements.  The physical origin of the perturbations is unclear,
although a variety of well-known accretion instabilities could
contribute \citep{fra92}.  Furthermore,
\citet{kin03} have recently shown that MHD turbulence can
cause accretion perturbations on sufficiently long time-scales for
a linear rms-flux relation to be produced by the inward propagation
and coupling of the perturbations. 

Finally, we note that a PP model origin for aperiodic variability in
XRBs and AGN implies that the variability can be used to probe
the location and structure of the X-ray emitting regions
close to the neutron star surface or black hole event horizon.  For
example, an extended emitting region acts as a low pass filter,
suppressing variations which originate within the emitting region relative to
slower variations which originate outside, for the simple reason that
an inward-propagating perturbation can only modulate emission within
the radius at which it is formed.  This simple picture can help to explain why
neutron star XRBs show stronger high-frequency noise (at frequencies $>10$~Hz) than
black hole systems \citep{sun00}, i.e. because their X-ray emission originates
predominantly on the neutron star surface, within the radius where the
highest frequency accretion flow perturbations are produced.
Similarly, in black hole systems the suppression of variability within a coronal emitting
region with a temperature gradient will lead to energy dependent
power-spectral shapes, which can be used to constrain the radial
emission structure of the corona (\citealt{kot01,zyc03}).

\section{Conclusions}
We have demonstrated that the aperiodic variability which carries the
linear rms-flux relation in the accreting millisecond pulsar
SAX~J1808.4-3658, is coupled to the 401~Hz pulsation in this source,
and hence the component of X-ray emission which contains the rms-flux relation
is produced at the magnetic caps of the
neutron star.  We conclude that the
aperiodic variability is produced by inward-propagating perturbations in the accretion
flow on to the neutron star, with the rms-flux relation produced by a
coupling of perturbations produced on different
time-scales.  By extension, this result suggests that the same
mechanism produces the rms-flux relations observed in black hole XRBs
and AGN, so that the aperiodic X-ray variability in all these diverse
systems is caused by perturbations in the accretion flow, and not by
flares due to magnetic reconnection in the corona.

\subsection*{Acknowledgments}
We wish to thank the anonymous referee for helpful comments.

\end{document}